\documentclass[flushrt,preprint,12pt]{aastex}
\tighten
\input psfig.sty
\usepackage{graphics}

\font\BFd=cmmib10 scaled 1200
\font\BFt=cmmib10 scaled 1200
\font\BFs=cmmib10
 
\def\bb#1{\relax
\ifmmode\mathchoice
{{\hbox{\BFd #1}}}{{\hbox{\BFt #1}}}
{{\hbox{\BFs #1}}}{{\hbox{\BFs #1}}}
\else \mbox{#1} \fi } 

\def\k{{\bb{k}}}
\def\xebar{\bar{x}_e}

\def\dxe{\delta_{x_e}}
\def\nebar{\bar{n}_e}
\def\nHbar{\bar{n}_H}

\def\rhobbar{\bar{\rho}_b}
\def\rhogbar{\bar{\rho}_\gamma}

\begin{document}
\author{Shwetabh Singh and Chung-Pei Ma}
\affil{Department of Physics and Astronomy, University of Pennsylvania,\\
  209 South 33rd Street, Philadelphia, PA~19104\\
	and\\
  Department of Astronomy, University of California at Berkeley,\\
  601 Campbell Hall, Berkeley, CA~94720\\
 \email{shwetabh@hep.upenn.edu;  cpma@astron.berkeley.edu}}
\title{Linear and Second-Order Evolution of Cosmic Baryon Perturbations \\
below $10^6$ Solar Masses}


\begin{abstract}
Studies of the growth of cosmic perturbations are typically focused on
galactic scales and above.  In this paper we investigate the evolution
of perturbations in baryons, photons, and dark matter for masses below
$10^6 M_\odot$ (or wavenumbers above 100 Mpc$^{-1}$).  Fluctuations on
these scales are of interest and importance because they grow to
become the earliest collapsed objects and provide the first light
sources in the so called dark ages.  We investigate both the linear
evolution and the second-order nonlinear effects arising from the coupling
of large-scale velocity fields to small-scale perturbations in the
baryon density and the electron ionization fraction.  We find that
this second order nonlinear coupling dominates the growth of
perturbations with masses $\la 10^{3} M_\odot$ immediately after
recombination, enhancing the baryon fluctuation amplitudes by a factor
of $\sim 5$, but the nonlinear effect does not persist at late times.
\end{abstract}

\section{Introduction}

Theories of structure formation are aimed at describing how gravity
affects the distribution of matter in the universe.  The currently
accepted paradigm is that gravitational instabilities amplify
fluctuations imprinted in the cosmological matter and radiation
density in the early universe, and these fluctuations eventually
evolve into the observed web of cosmic structures.  By assuming these
fluctuations to be small, a linear perturbation theory can be
formulated and solved.  The linear theory for a perturbed
Friedmann-Robertson-Walker metric was developed by Lifshitz (1946).
The theory and subsequent modifications have been applied to a large
number of cosmological problems and are described in detail in several
textbooks (e.g., Weinberg 1972; Peebles 1980; Peacock 1999).

Most calculations and analyses based on the cosmological linear
perturbation theory are focused on galactic scales or above at masses
of $M\ga 10^9 M_\odot$, or wavenumbers of $k \la 10$ Mpc$^{-1}$.  This
is likely due to the fact that quantities such as the mass fluctuation
power spectrum can be determined observationally only on these large
scales, and that the computations become expensive at high wavenumbers
due to the rapid oscillations in the perturbed fields.  The theory,
however, is equally valid for small scales, and the evolution of $M
\la 10^6 M_\odot$ perturbations is key to the understanding of the
formation of the first generation objects in the high redshift
universe.  A detailed study of perturbation modes up to $k\sim 1000$
Mpc$^{-1}$ has found interesting features in the linear baryon field
(Yamamoto, Sugiyama, \& Sato 1998).  For example, the tight coupling
between baryons and photons, which is assumed until the epoch of
recombination in most calculations, breaks down well before
recombination on these very small scales.  This allows the baryon
density field to grow substantially by recombination under the
influence of the cold dark matter (CDM).  This behavior is contrary to
the common notion that the baryon fluctuations are wiped out by photon
diffusion or Silk damping (Silk 1968) on small scales.  After
recombination, on the other hand, perturbation modes with $k\ga 300$
Mpc$^{-1}$ are below the Jeans mass and exhibit oscillatory behavior,
in contrast to the lower-$k$ modes which grow via Jeans instability
immediately after recombination.  This retards the baryon growth until
the mode becomes Jeans unstable again at a later time.

Beyond the linear order, there have been recent discussions regarding
nontrivial second-order effects that can enhance the linear growth of
fluctuations on very small scales during recombination (Shaviv 1998;
Liu et al. 2001).  This nonlinear effect arises from the coupling of
the large scale velocity fields to the perturbations in the baryon
density $\delta_b$ and the electron ionization fraction $\delta_{x_e}$
on small scales.  An amplification of $\sim 10^4$ in $\delta_b$ has
been reported (Shaviv 1998), although a more in-depth analysis (Liu et
al. 2001) has found that the dramatic increase in the baryon amplitude
is a result of neglecting the electron-photon diffusion term, which
when included, would reduce the nonlinear enhancement to a factor of
$\sim 10$.  More specifically, Shaviv (1998) showed that the time
scale governing the growth of small scale perturbations in the baryon
density is $\sim \tau_{R}^2/\tau_{0}$, where $\tau_{R}$ is the
characteristic time scale for the second-order terms and $\tau_{0}$ is
the time scale for acoustic oscillations in the baryon fluid.  He took
the time scale $\tau_D$ for momentum transfer between electrons and
photons due to Thomson scattering to be much larger than
$\tau_{R}^2/\tau_{0}$.  Liu et al. (2001) pointed out that this
assumption is invalid during recombination since $\tau_D\sim
\tau_R^2/\tau_0$ and $\tau_D$ cannot be ignored.  In addition, Shaviv
(1998) used the Saha equation to evaluate $\delta_{x_e}$ which is
needed to calculate $\tau_{R}$.  Liu et al. (2001) instead used a rate
equation for $\delta_{x_e}$ because the assumption of thermal
equilibrium required for the validity of the Saha equation does not
hold during recombination.  They then assumed a wave form solution for
the baryon density field $\delta_b \propto \exp{(\omega \tau-ikx)}$,
and estimated the growth of $\delta_b$ from the real part of $\omega$,
which they obtained from the dispersion relation given by the baryon
fluid equation.

In this paper, we investigate this small-scale nonlinear phenomenon in
its full extent by using the complete cosmological perturbation
theory.  Our work differs from Liu et al. (2001) in that instead of
tracking only the real part of $\delta_b$ from the dispersion
relation, we calculate the evolution of the perturbations in all
particle species (baryons, photons, CDM, and neutrinos) at high
wavenumbers $k > 1000$ Mpc$^{-1}$ by solving the complete set of
Einstein and Boltzmann equations consistently.  We use the linear
Boltzmann code in the COSMIC package (Ma \& Bertschinger 1995;
Bertschinger 1995) and incorporate the aforementioned nonlinear
effects by adding second-order contributions to the photon-electron
momentum transfer terms due to Thomson scattering.  For consistency,
we include these terms in the evolution equations for photons as well
as baryons.  This approach allows us to take into account possible
nonlinear coupling effects and to calculate the growth, decay, and
oscillations in all particle species accurately.

In \S~2.1 we summarize the key portions of the linear cosmological
perturbation theory.  In \S~2.2 we introduce the nonlinear coupling
terms in the baryon and photon equations as well as the equations for
the perturbed ionization fraction.  Numerical results for the linear
and the nonlinear evolution of four representative high-$k$ modes,
$k=1000$, 2500, 5000, and 10000 Mpc$^{-1}$, are presented in \S~3.
The corresponding masses are $\sim 1000$, 64, 8, and 1 $M_\odot$,
respectively.  \S~4 includes a discussion of the physical origin of
this phenomenon and the reason why a second order term could dominate
the growth of perturbations at the era of recombination.

\section{Cosmological Perturbation Theory}

\subsection{Linear Perturbations}

The discussion here is restricted to the scalar modes of metric
perturbations.  We will work in the conformal Newtonian gauge
\begin{equation}
   ds^{2}=a^2(\tau)[-(1+2\psi)d\tau^{2} + (1-2\phi)dx^2]\label{ec1}\,,
\end{equation}
where $\phi$ and $\psi$ are the two scalar potentials that
characterize the perturbations, $\tau$ is the conformal time, $x$ are
the spatial dimensions, and $a$ is the scale factor.  The scalar
potential $\psi$ has a simple physical interpretation of being the
gravitational potential in the Newtonian limit, and $\phi\approx\psi$
in the absence of massive neutrinos.

We solve the full set of linear evolution equations for small
perturbations in the metric and the phase-space distributions of CDM,
photons, baryons, and neutrinos.  The full theory in this gauge is
described in detail in Mukhanov et al. (1992) and Ma and Bertschinger
(1995); here we write out only the lowest two velocity moments of the
phase space distribution for CDM (subscript $c$) , baryons ($b$), and
photons ($\gamma$), which are the key quantities in this study.  In
$k$-space, we have
\begin{eqnarray}
   \dot {\delta_{c}} &=& -\theta_{c} + 3\dot {\phi} \,, \nonumber\\
   \dot {\theta_{c}} &=& -\frac{\dot {a}}{a}\theta_{c} + k^{2}\psi \,,
	\nonumber\\
   \dot {\delta_{b}} &=& -\theta_{b} + 3\dot {\phi} \,,\nonumber\\
   \dot {\theta_{b}} &=& -\frac{\dot {a}}{a}\theta_{b} + 
        c_{s}^2k^2\delta_{b} + k^{2}\psi + 
        \frac{4 \rhogbar}{3 \rhobbar} a \nebar
        \sigma_{T}(\theta_{\gamma}-\theta_{b}) \,, \label{linear} \\
  \dot {\delta_{\gamma}} &=& -\frac{4}{3}\theta_{\gamma} + 
        4\dot {\phi} \,, \nonumber\\
   \dot {\theta_{\gamma}} &=& k^2(\frac{1}{4}\delta_{\gamma}
	-\sigma_{\gamma})   + k^2\psi 
       + a \nebar \sigma_{T}(\theta_b-\theta_\gamma) \,, \nonumber       
\label{master}
\end{eqnarray}
where $\delta$ is the density fluctuation, $\theta=i\k \bb{v}$ is
divergence of the fluid velocity, and $\sigma_\gamma$ is the sheer
stress of the photon field which is coupled to higher moments not
written down here.  On the right hand side, $\sigma_T$ is the Thomson
scattering cross section, $c_s$ is the baryon sound speed, $\nebar$ is
the mean electron number density, and $\rhogbar$ and $\rhobbar$ are
the mean photon and baryon energy densities.  Dots denote derivatives
with respect to the conformal time, and the densities and wavenumbers
are all comoving.  Massive neutrinos have no direct coupling to the
nonlinear terms considered here, so we do not include them.

Additional equations are needed to calculate $\nebar$.  We define the
mean ionization fraction $\xebar$ as
\begin{equation}
	\xebar \equiv {\nebar \over  \nHbar }
        = { \nebar m_p \over (1-y) \rhobbar } \,,
\label{Xe}
\end{equation}
where $\nHbar$ is the mean hydrogen number density, $y$ is the
primordial helium mass fraction, and $m_p$ is the proton mass.  The
value $\xebar=1$ corresponds to complete hydrogen ionization and
neutral helium, and $\xebar$ can exceed unity when helium is ionized
at an earlier time.  We choose this convention for convenience because
for the redshift of interest to this paper, hydrogen is the dominant
source of free electrons.  For times much before recombination,
$\xebar$ obeys the equilibrium Saha equation to a good approximation,
\begin{equation}
    { \xebar^2 \over 1-\xebar} = {1\over \nHbar} 
	\left( \frac{m_e k_B T_b}{2 \pi \hbar^2} \right)^{3/2} 
	e^{-13.6\, {\rm eV}/k_B T_b} \,.
\end{equation}
Here $T_{b}$ is the baryon temperature, $k_{B}$ is the Boltzmann
constant, $m_{e}$ is the electron mass, and $\hbar$ is the Planck's
constant.  During recombination the rapidly declining free electron
density leads to a breakdown of ionization equilibrium, and one must
integrate the appropriate kinetic equations.  We use the recombination
rate equation (Peebles 1968; Spitzer 1978)
\begin{equation}
   \frac{d\xebar}{d\tau}=a C_r
	[\beta(T_{b})(1-\xebar) - \nHbar \alpha^{(2)}(T_b) \xebar^2]\,,
\label{rate0}
\end{equation}
where the collisional ionization rate from the ground state is
\begin{equation}
    \beta(T_b) = \left( {m_e k_B T_b \over 2\pi\hbar^2} \right)^{3/2}\,
	e^{-13.6\, {\rm eV}/ k_B T_b} \alpha^{(2)}(T_b) \,,
\end{equation}
and the recombination rate to the excited states is
\begin{equation}
   \alpha^{(2)}=\frac{64 \pi}{(27 \pi)^{1/2}} \frac{e^4}{m_e^2 c^3} 
   \left( {k_B T_b \over 13.6\, {\rm eV}} \right)^{-1/2} \phi_2 (T_b)\,, 
	\quad
    \phi_2(T_b)=0.448 \ln \left({ 13.6\, {\rm eV}\over k_B T_b}\right) \,.
\end{equation}
Here $e$ is the electron charge and $c$ is the speed of light. The net
recombination rate to the ground state is reduced by the fact that an
atom in the $n=2$ level may be ionized before it decays to the ground
state.  The Peebles' reduction factor, $C_r$, is the ratio of the net
decay rate to the sum of the decay and ionization rates from the $n=2$
level,
\begin{equation}
   C_r = \frac{\Lambda_{\alpha} + \Lambda_{2s\rightarrow 1s}}
	{\Lambda_{\alpha} + \Lambda_{2s\rightarrow 1s} 
	+ \beta^{(2)}(T_{b})} 
\end{equation}
where
\begin{equation}
   \beta^{(2)}(T_b) = \beta(T_b) \, e^{hc/k_B T_b \lambda_\alpha} \,,
	\quad
   \Lambda_\alpha = \frac{8 \pi \dot a}{a^2 n_{1s} \lambda_{\alpha}^3} \,,
	\quad
   \lambda_{\alpha}=1.216 \times 10^{-5} \,,
\label{ec26}
\end{equation}
$\Lambda_{2s\rightarrow 1s}$ is the rate at which net recombination
occurs through two-photon decay from the 2$s$ level and is equal to
8.227 s$^{-1}$, and $n_{1s}$ is the number density of hydrogen atoms
in the 1$s$ state.

\subsection{Second-Order Perturbations}

A complete list of second order nonlinear contributions in
perturbation theory includes many terms, but the dominant contribution
comes from the terms of order $\delta_{b} v_{b}$ (e.g., Hu, Scott, and
Silk 1994; Dodelson and Jubas 1995).  The other second order terms are
proportional to $v_b^{2}$, which are suppressed because $v_b \sim
\delta_{b}/ k\tau \ll \delta_b$ for the scales of interest here with
$k\tau \gg 1$.  It will be demonstrated later in the section that the
perturbed ionization fraction $\delta_{x_e}$ is of the same order of
magnitude as $\delta_b$ for the scales of interest.  We will therefore
include perturbations to both the ionization fraction and the baryon
density as the terms giving rise to second order effects in the
evolution equations.

It is easier to determine such second-order terms in real space.  The
relevant first-order equations to be modified are the $\dot\theta_b$
and $\dot\theta_\gamma$ equations in (\ref{master}), which in real
space take the form
\begin{eqnarray}
      \dot {v_b} &=& -\frac{\dot {a}}{a} v_b - c_{s}^2 \nabla\delta_{b} - 
	\nabla\psi + {4 \rhogbar \over 3 \rhobbar} a \nebar \sigma_{T} 
	(v_{\gamma}-v_{b})\,, \nonumber\\ 		
      \dot {v_{\gamma}} &=& 
	-\nabla\left( \frac{1}{4}\delta_{\gamma} - \sigma_{\gamma} \right)
	-\nabla\psi + a \nebar \sigma_{T}(v_{b}-v_{\gamma}) \,. 	
\label{V_bg0}
\end{eqnarray}
Allowing for spatial fluctuations in the ionization fraction $x_e$,
\begin{equation}
   x_e = \xebar (1 + \dxe) \,,
\end{equation}
the perturbations in the electron density then contain two terms
arising from inhomogeneities in the baryon density and the ionization
fraction, respectively:
\begin{equation}
   n_e = \nebar (1 + \delta_{n_e})=\nebar (1+\delta_b + \dxe) \,.
\end{equation}
We obtain the second-order contributions to the Thomson scattering
terms in equation~(\ref{V_bg0}) by replacing $\nebar$ with $n_e$,
$\rhobbar$ with $\rho_b=\rhobbar(1+\delta_b)$, and $\rhogbar$ with
$\rho_\gamma=\rhogbar(1+\delta_\gamma)$:
\begin{eqnarray}
      \dot {v_b} &=& -\frac{\dot {a}}{a} v_b - c_{s}^2 \nabla\delta_{b} - 
	\nabla\psi + {4 \rhogbar \over 3 \rhobbar} a \nebar \sigma_{T} 
      (1+\dxe+\delta_\gamma) (v_{\gamma}-v_{b})\,, \nonumber\\
      \dot {v_{\gamma}} &=& 
     -\nabla\left( \frac{1}{4}\delta_{\gamma} - \sigma_{\gamma} \right)
     -\nabla\psi + a \nebar \sigma_{T} (1+\dxe+\delta_b) 
     (v_{b}-v_{\gamma})\,.
\label{V_b}
\end{eqnarray}
We will drop the second-order term containing $\delta_{\gamma}$ in the
$\dot{v}_b$ equation because on small scales, $\delta_{\gamma} \ll
\delta_{x_e} \sim \delta_{b}$ due to photon diffusion damping.
Transforming back to $k$ space, the nonlinear terms of ${\cal
O}(\delta\, v)$ become convolutions, and we have
\begin{eqnarray}
      \dot {\theta_b} & = & -\frac{\dot {a}}{a}\theta_{b} 
 	+ c_{s}^2k^2\delta_{b} + k^{2}\psi 
        + {4 \rhogbar \over 3 \rhobbar} a \nebar \sigma_{T}
	(\theta_{\gamma}-\theta_{b}) \nonumber\\ 
      & & + {4 \rhogbar \over 3 \rhobbar} a \nebar\sigma_{T} \, k
      \int d^3 k'\, [v_{\gamma}(k') -v_{b}(k')]\, \dxe(|\k-\k'|) \, 
	 \nonumber\\ 	
      \dot {\theta_{\gamma}} & = & k^2 \left( \frac{1}{4}\delta_{\gamma}
        -\sigma_{\gamma}\right) + k^2\psi + a \nebar  
      \sigma_{T}(\theta_{b}-\theta_{\gamma}) \nonumber\\
      & & +  a \nebar \sigma_{T} \, k
	\int d^3 k' \, [v_{b}(k')-v_{\gamma}(k')]\, [\dxe(|\k-\k'|) +
	\delta_{b}(|\k-\k'|)]  \,,  
\label{theta}
\end{eqnarray}
We also need an evolution equation for the ionization fraction
fluctuation $\dxe$.  It can be obtained by perturbing the zeroth-order
ionization rate equation (\ref{rate0}) (Liu et al. 2001):
\begin{equation}
   {d\, \dxe \over d \tau}  =  -a C_r \left[ \beta \,\dxe + 
	\nHbar \alpha^{(2)} \xebar \left(\delta_b + 2 \dxe \right) \right]
  + \left( {\delta C_r \over C_r} - \dxe \right) {1\over \xebar}
    { d \xebar \over d\tau} \,,
\label{rate1}
\end{equation}
where
\begin{equation}
  \frac{\delta C_r}{C_r}=-\frac{\Lambda_\alpha n_{1s}
      \beta^{(2)}\, \nHbar [ (1-\xebar) \delta_b - \xebar \dxe]} 
   {[\Lambda_\alpha n_{1s} + \Lambda_{2s \rightarrow 1s} \nHbar (1-\xebar)]
    [\Lambda_\alpha n_{1s} + (\Lambda_{2s\rightarrow 1s} + \beta^{(2)})
	\nHbar (1-\xebar) ]} \,.
\end{equation}

We have assumed that $\delta_{T_{b}}=0$, i.e., there are no
perturbations induced in the baryon temperature.  We verify this
claim by including the perturbed evolution equation for $T_{b}$ (Ma and
Bertschinger 1995) in our calculations:
\begin{equation}
    \frac{d (\delta_{T_{b}})}{d \tau}=\delta_{T_{b}} 
\left[\frac{\dot{\bar{T_{b}}}}{\bar{T_{b}}} 
- \frac{2 \dot{a}}{a} \right] +\frac{8 \bar{\rho_{\gamma}} \mu}{3 \bar{T_{b}} 
m_{e} 
\bar{\rho_{b}}} a \bar{n_{e}} \sigma_{T} 
\left[\delta_{x_e}(T_{\gamma}-\bar{T_{b}}) 
-\bar{T_{b}} \delta_{T_{b}} \right]\,.
\label{deltatb}
\end{equation}
\noindent Here $\mu$ is the mean molecular weight (including free electrons and 
all ions of Hydrogen and Helium) and $T_{b}=\bar{T_{b}}(1+\delta_{T_{b}})$.We 
find that the value for $\delta_{T_{b}}$ is always at least two
orders of magnitude smaller than $\delta_{x_e}$ because the source
term $(T_{\gamma}-T_{b})$ for $\delta_{T_{b}}$ is negligible until well
into the recombination era.  Even after recombination the growth is
not rapid enough to affect the evolution of $\delta_{x_e}$.

We note that the perturbation to the ionization fraction $\dxe$ arises
naturally from the first-order cosmological equations and is not an
ad hoc term.  This perturbation is ignored in standard linear
calculations because its effects on the photon and baryon evolution
enter only at second order, as shown by equations~(\ref{V_b}) and
(\ref{theta}).  We show in the next section that numerical results
from the linear theory indeed give comparable amplitudes for $\dxe$
and $\delta_b$.  We are thus justified in keeping both the $\delta_b
v$ and $\dxe v$ terms in our second order analysis.

\section{Numerical Results}

In this section we present results from numerical integration of the
linear and second-order equations in \S~2.  The calculations are
performed with the full Boltzmann code for the conformal Newtonian
gauge in the COSMIC package (Ma \& Bertschinger 1995; Bertschinger
1995).  For the second-order results, modifications are made to this
code to include the nonlinear terms in equation~(\ref{theta}) and the
perturbed rate equation (\ref{rate1}).  The matter density parameter
is taken to be $\Omega_m=0.35$ with $\Omega_c=0.30$ in CDM and
$\Omega_b=0.05$ in baryons.  The cosmological constant and Hubble
parameter are $\Omega_\Lambda=0.65$ and $h=0.75$.  Three species of
massless neutrinos and a helium fraction $y=0.24$ are assumed.  The
initial conditions are adiabatic, Gaussian fluctuations with a
spectral index of $n=1$.  All results are normalized to COBE.

\subsection{Linear Evolution}

Figure 1 illustrates the time evolution of the linear baryon density
field $\delta_b$ and CDM density field $\delta_c$ for wavenumbers
$k=1000$, 2500, 5000, and $10^4$ Mpc$^{-1}$.  The corresponding masses
are $\sim 1000$, 64, 8, and 1 $M_\odot$, respectively.  These scales
are much smaller than the typically studied scales of $k\la 10$
Mpc$^{-1}$ or $M\ga 10^9 M_\odot$.  We note two interesting features
that are absent in the more familiar behavior of $\delta_b$ at lower
$k$.  First, even though photon diffusion damping wipes out $\delta_b$
very rapidly at $z\sim 10^6$ to $10^5$, $\delta_b$ at such high $k$ is
able to grow before recombination due to the break down of
photon-electron tight coupling on small scales.  This rapid
regenerative growth is characterized by a $(1+z)^{-7/2}$ dependence and
can be understood in terms of the balance achieved between the
radiation drag force and the gravitational force (Yamamoto et
al. 1998).  As a result, $\delta_b$ grows by at least 5 orders of
magnitude before recombination ends at $z\sim 1000$.  We also note a
short period during recombination when $\delta_b$ grows even more
rapidly.  This is because gravity is rapidly becoming the dominant
force over the decreasing radiation drag force.

The second feature special to these high $k$ modes is that they
undergo a second Jeans length crossing after recombination.  It is
well known that the baryonic Jeans wavenumber $k_J =\lambda_J^{-1}
\sim c_s^{-1} ( 4\pi a^2 G\rho)^{1/2}$ increases rapidly during
recombination, reaching $k_{J,rec} = 900 (\Omega_m h^2)^{1/2} {\rm
Mpc}^{-1} \approx 370$ Mpc$^{-1}$ shortly after recombination for our
model (Yamamoto et al. 1998).  Perturbations with $k>k_{J,rec}$, which
were Jeans unstable before recombination, become stable after
recombination and exhibit oscillations due to the baryon thermal
pressure.  The growth of $\delta_b$ is therefore slowed down as can
been seen in Figure~1.  As we will show below, it is during this era
when the small scale perturbations are relatively constant that the
second order nonlinear term becomes important.  When the Jeans
wavenumber grows to the scale of a given $k$ mode eventually, the
growth of $\delta_b$ picks up again.

Figure~2 shows the linear power spectrum for the velocity difference
$(v_b-v_\gamma)$ between the baryons and photons shortly after
recombination.  The Thomson scattering term is proportional to this
quantity, which serves as the source term for the second-order effect
to be discussed in the next section.

\subsection{Nonlinear Evolution}

We now solve the Einstein and Boltzmann equations taking into account
the second-order terms in equation~(\ref{theta}).  Two new features
must be handled: the evolution of the perturbed ionization fraction
$\dxe$ given by equation~(\ref{rate1}), and the $k$-convolution in
equation~(\ref{theta}).

To compute $\dxe$, we add equation~(\ref{rate1}) directly in the
COSMIC Boltzmann code.  We note that even though $\dxe$ is a
first-order quantity and is of comparable amplitude as $\delta_b$,
equation~(\ref{rate1}) was not included in the original linear COSMIC
code because $\dxe$ contributes to the photon and baryon evolution
equations only at second order.  Figure~3 illustrates our numerical
result for the time evolution of $\dxe$ for the $k=5000$ Mpc$^{-1}$
mode.  The dashed curve is for $\dxe$ computed from the linear theory.
It indeed has a similar amplitude as $\delta_b$ in Figure~1.
Including the second-order terms in the calculations enhance the
amplitude of $\dxe$ (solid curve), which peaks at a redshift of $\sim
1000$.

To handle the convolution terms in equation~(\ref{theta}), we use
the property that the velocity difference $(v_\gamma-v_b)$ in the
integrand has significant contributions only from $k'=0.01$ to 1
Mpc$^{-1}$, as shown in Figure~2.  This is much smaller than $k\ga
1000$ Mpc$^{-1}$ investigated in this paper.  This allows us to take
out $\dxe$ and $\delta_b$ from the convolutions and approximate the
velocity integral as
\begin{equation}
      \int d^3 k'\, [v_{\gamma}(k') -v_{b}(k')]\, \dxe(|\k-\k'|) 
	\approx \sigma_v \, \dxe(k) \,,
\end{equation}
where
\begin{equation}
   \sigma_v^2 = 4\pi \int d\ln k' \, k'^3 [v_\gamma (k')-v_b(k')]^2 \,.
\end{equation}
Figure~4 shows $\sigma_v$ as a function of redshift.  Initially
$\sigma_v$ is small because the tight coupling between baryons and
photons keeps $v_\gamma\approx v_b$.  As recombination proceeds and
the tight coupling breaks down, the baryons fall into the CDM
potential well, giving rise to the rapidly growing $\sigma_v$ during
$z\sim 1500$ to 1000.  For comparison, the long-dashed curve shows the
baryon velocity alone.  The agreement between the two curves shows
that the photon velocity field $v_\gamma$ is negligible after
recombination.  We approximate the time dependence of $\sigma_v$ with
the fitting formula
\begin{equation}
	\log_{10} \sigma_v=-3.57+9.56\times 10^{-4} z 
	- 3.3\times 10^{-6} z^2 +1.126\times 10^{-9} z^3
\end{equation}
shown as the short-dashed curve in Figure~4.  This formula is used in
the numerical integration for computational efficiency.

The results of our nonlinear calculations are illustrated in
Figures~5-7.  Figure~5 shows the time evolution of the linear vs.
nonlinear baryon density field $\delta_b$ for a single mode $k=5000$
Mpc$^{-1}$ of $M\approx 10 M_\odot$.  The nonlinear $\delta_b$ is
calculated from the full Boltzmann code with the second order
convolution terms.  The conformal time $\tau$ is used for the
horizontal axis to illustrate the post-recombination Jeans
oscillations which are special to these high-$k$ modes (see also
discussions in \S~3.1).  It can be seen that $\delta_b$ begins to
oscillate immediately after recombination at $\tau \approx 200$ Mpc
due to the sharp decrease in the Jeans length.  The time period of
these oscillations is roughly $2 \pi / k c_s$, where the baryon sound
speed is related to the baryon temperature by $c_s^2=k_B T_b(1-d\ln
T_b/d\ln a/3)$.  For $k=5000$ Mpc$^{-1}$, the baryons have $T_b\sim
2500$ K at time $\tau \sim 200$ Mpc, giving $c_s \sim 1.5 \times
10^{-5}$ and $\Delta\tau \sim 100$ Mpc, which agrees well with the
periods of oscillations shown in Figure~5.  Figure~6 shows the ratio
of nonlinear and linear $\delta_b$ for four wavenumbers: $k=1000$,
2500, 5000, and 10000 Mpc$^{-1}$.  Figure~7 shows the same thing for
the photon density field $\delta_\gamma$, which has a similar behavior
as the baryons.

We find that the inclusion of the second-order terms has two main
effects on the baryons: an increase in the amplitude of the Jeans
oscillations, and a slight shift in the sound speed and hence the
oscillation period.  Both effects are clearly seen in the figures.
The nonlinear enhancement reaches a maximum of a factor of several
right after recombination.  It then decreases as the mean ionization
fluctuation $x_e$ decreases.  The slight shift in the oscillation
period is caused by the inclusion of perturbations to the ionization
fraction.  We have effectively introduced another term in the
evolution equation for $\theta_{b}$, which together with the linear
terms determine the period of the oscillations.  The nonlinear
coupling thus changes the time period of these oscillations to $\sim$
2$\pi$/($kc_{s}$ + nonlinear effects).  This slight shift in the
effective sound speed is the reason for the oscillations in Figures~6
and 7.  It is also worthwhile to note that the maximum amplitude
achieved by the ratio of the baryonic perturbations in the nonlinear
theory compared to the linear calculation does not follow a simple
relationship as a function of $k$.  The amplitude of the ratio, for
example, is larger for $k=5000$ Mpc$^{-1}$ than for $k=10000$
Mpc$^{-1}$.  This is because the effect of the nonlinear term becomes
less important as the frequency of the Jeans oscillations increases.
For the particular cosmological model under consideration this occurs
at a scale between $k=5000$ and 10000 Mpc$^{-1}$.

\section{Discussion}

Figures~6 and 7 show that the second order terms arising from the
coupling of the large scale baryon velocity fields to the perturbation
in the electron ionization fraction on small scales dominate the
evolution equation for the baryon density perturbations $\delta_b$ at
$z\sim 1000$.  This second order effect dies off at lower $z$,
however, and it does not push the small scale baryon fluctuations into
the nonlinear regime at a significantly earlier time as was suggested
by Shaviv (1998).

It is nevertheless interesting that for modes with $k > 1000$
Mpc$^{-1}$, a second-order term can significantly alter the growth
predicted by the linear theory right after recombination.  How does
any second-order term at $z\sim 1000$ lead to an amplification of a
factor of $\sim 5$ in the linear baryon fluctuations as we have found?
To understand this, we plot in Figure 8 the time evolution of all the
terms on the right-hand-side of the $\dot\theta_b$ equation from
(\ref{theta}): $-\dot{a} \theta_b/a$ (short-dashed), $c_s^2 k^2
\delta_b$ (dotted), $(4 \rhogbar/ 3 \rhobbar) a \nebar \sigma_T
(\theta_\gamma-\theta_b)$ (long dashed), and the second-order term
$\sim v_b \dxe$ (dot-dashed).  Since the potential term $k^2 \psi$ is
nearly a constant of time, it is convenient to normalize other terms
to $k^2 \psi$.  To contrast the scale dependence of the nonlinear
effect, we show two modes: $k=250$ and 5000 Mpc$^{-1}$.  As one can
see, the nonlinear effect is important only for the higher-$k$ mode.
The nonlinear term is unimportant for the $k=250$ Mpc$^{-1}$ mode
because the pressure term $k^2 c_s^2 \delta_b$ is small on these
scales and can not prevent the perturbations from growing under the
influence of gravity.  For the $k=5000$ Mpc$^{-1}$ mode, however, the
pressure term $k^2 c_s^2 \delta_b$ keeps the perturbations in the
terminal velocity stage for a longer period of time (see Yamamoto et
al. 1998).  This leads to a near cancellation of all linear terms
governing the growth of $\theta_b$.  As a result, the second-order
term gains importance and through the nonlinear coupling becomes the
dominant term shortly after recombination.  The effect of the coupling
on the perturbed ionization fraction is shown in Figure~3, where the
nonlinear coupling can give rise to an order of magnitude change in
the value of $\dxe$.  At late times the second-order term becomes
negligible as the free electron density $\nebar$ becomes very small.
We note that the Hubble drag term $-\dot{a} \theta_b / a$ is
unimportant throughout because the scales of the perturbations under
consideration are so small that they are not influenced by the large
scale expansion of the universe.

Our approach and results differ from the previous work on this subject
(Liu et al. 2001) in a number of ways.  We have solved the full set of
Einstein and Boltzmann equations for the metric perturbations and the
perturbation field in CDM, baryons, photons, and neutrinos.  This
approach has allowed us to calculate accurately the phase space
distributions of both baryons and photons.  For the baryons, we find
that the maximum amplitude reached by the nonlinear to linear ratio,
$\delta_{b}$(Nonlinear)/$\delta_{b}$(Linear), is similar to that of
Liu et al. (2001) for $k \la 5000$ Mpc$^{-1}$, but our results show an
oscillating ratio because the addition of the nonlinear coupling
causes a change in the effective speed of sound.  This effect was
unnoticed in the work of Liu et al. (2001).  In addition, their
analysis finds a monotonic increase in the amplitude of the ratio as a
function of $k$.  We find instead that the maximum amplitude reached
by the ratio for $k=10000$ Mpc$^{-1}$ is less than that for $k=5000$
Mpc$^{-1}$.  This is again because their analysis was unable to track
the Jeans oscillations after recombination accurately.

Our calculation has also included the nonlinear coupling terms in the
evolution of the photon density perturbations.  As
equations~(\ref{V_b}) and (\ref{theta}) show, two comparable nonlinear
terms contribute to the photon velocity field: $\dxe
(v_{b}-v_{\gamma})$ and $\delta_b (v_{b}-v_{\gamma})$.  We have shown
that the $\delta_{x_e} (v_{b}-v_{\gamma})$ term results in an initial
enhancement of the amplitude for the baryons.  Momentum conservation
in Thomson scattering implies that this nonlinear term would tend to
suppress the amplitude obtained from the linear calculation for the
photons.  It turns out that the $\delta_{b} (v_{b}-v_{\gamma})$ term
is generally larger and it acts in an opposite sense as the
$\delta_{x_e} (v_{b}-v_{\gamma})$ term.  The end result is an
enhancement in $\delta_\gamma$ as well, as shown in Figure~7.  It may
be an interesting future study to compute directly the angular power
spectrum for the temperature fluctuations in the cosmic microwave
background, but we expect the angular scales associated with the $k
\ga 1000$ Mpc$^{-1}$ modes to be at multipoles of $l \ga 10^6$, below
the scales probed by the current generation of experiments.

\acknowledgements
 
We thank Ue-Li Pen, Naoshi Sugiyama, and Thanu Padmanabhan for useful
discussions.  C.-P. M. acknowledges support of an Alfred P. Sloan
Foundation Fellowship, a Cottrell Scholars Award from the Research
Corporation, a Penn Research Foundation Award, and NSF grant AST
9973461.


\clearpage
\begin{figure*}
\begin{tabular}{c}
\psfig{file=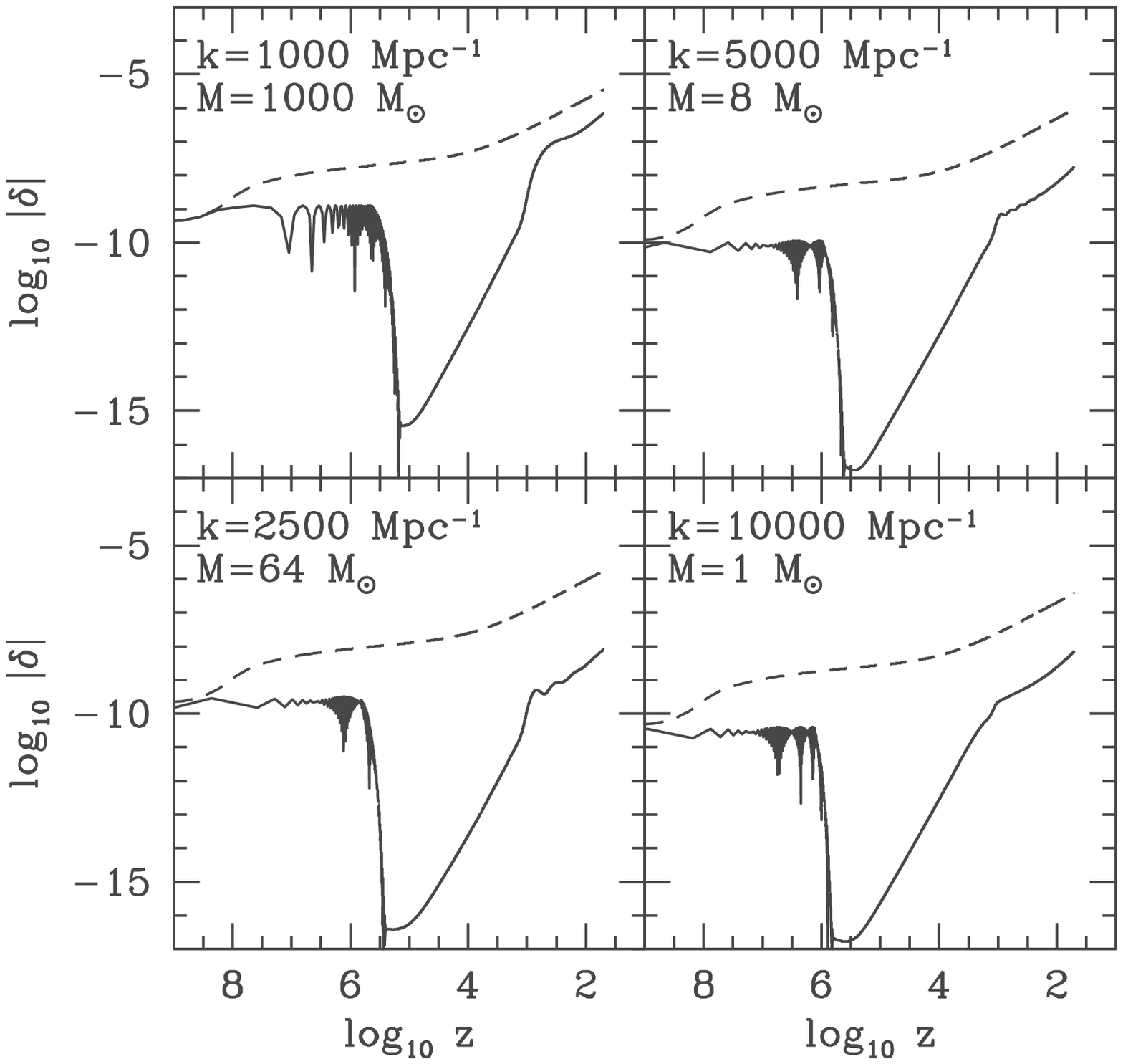,width=6in} 
\end{tabular}
\figurenum{1}
\caption{ Time evolution of the linear baryon (lower curve) and cold
dark matter (upper curve) density field on very small scales: $k=
1000$, 2500, 5000, and 10000 Mpc$^{-1}$.  The results are computed in
the conformal Newtonian gauge using the COSMIC Boltzmann code.  The
cosmological parameters are $\Omega_c=0.30, \Omega_b=0.05,
\Omega_\Lambda=0.65, h=0.75$, and the fluctuation amplitudes are
normalized to COBE.  Photon diffusion damping results in the sharp
drop in the baryon amplitude at $z\sim 10^6$ to $10^5$.  This is
followed by a power law growth when the photon-baryon tight coupling
becomes ineffective.  The small oscillations after recombination at
$z\sim 1000$ in the high-$k$ modes are a result of the second Jeans
length crossing (see text).}
\end{figure*}

\clearpage
\begin{figure*}
\begin{tabular}{c}
\psfig{file=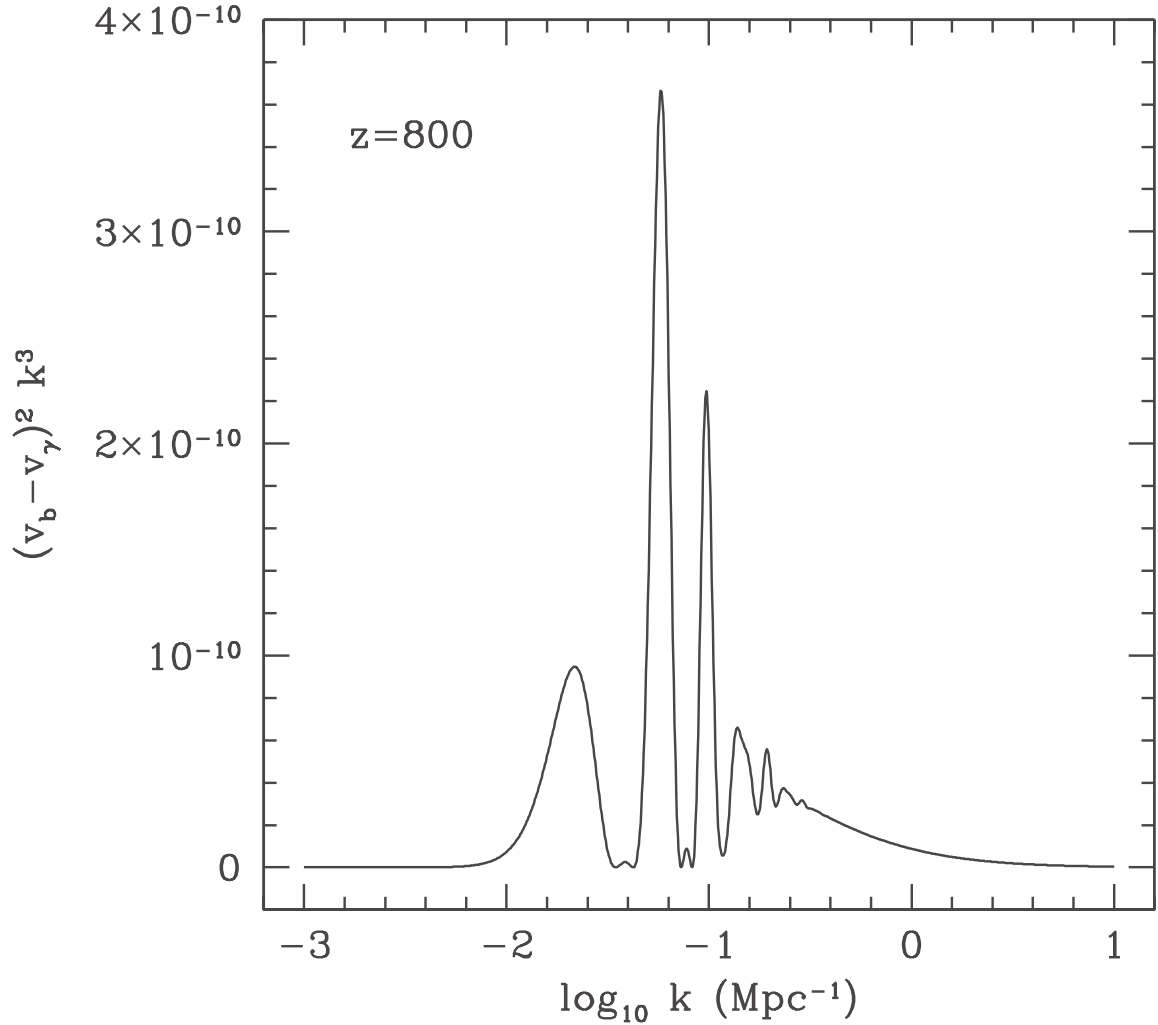,width=6in} 
\end{tabular}
\figurenum{2}
\caption{Linear power spectrum for the velocity difference between baryons
and photons at redshift $z=800$.  The dominant contributions
come from the modes in the range of $k\sim 0.01$ to 1 Mpc$^{-1}$,
which is much below the modes of interest in Fig.~1.  This feature
allows us to simplify the second-order velocity convolution terms in
eq.~(\ref{theta}).}
\end{figure*}

\clearpage
\begin{figure*}
\begin{tabular}{c}
\psfig{file=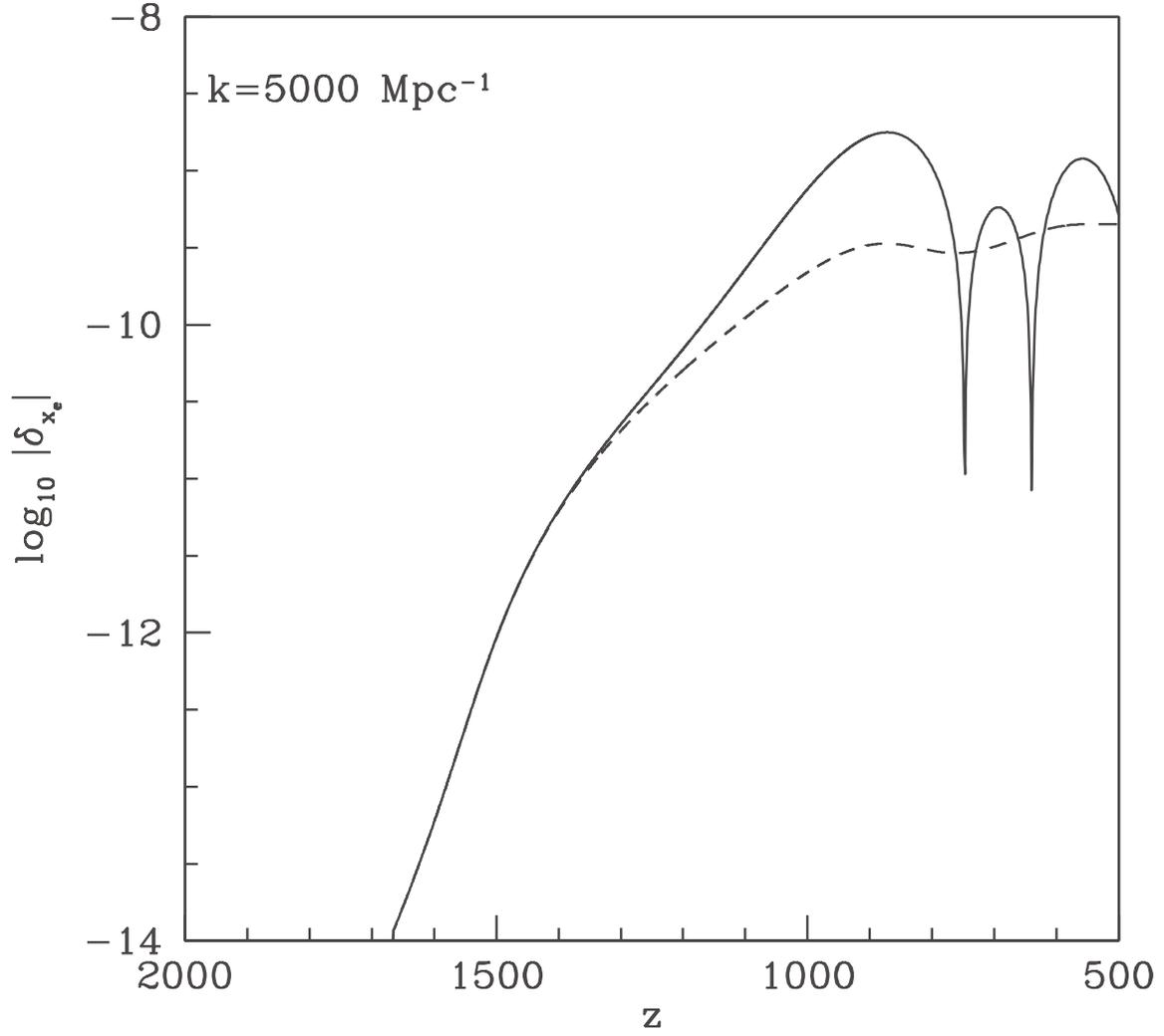,width=6in} 
\end{tabular}
\figurenum{3}
\caption{Perturbation to the ionization fraction, $\dxe$, as a
function of redshift, computed without (dashed curve) and with (solid
curve) the second-order terms in eq.~(\ref{theta}) for the velocity
field $\dot\theta$.  The peak occurs $z \sim 1000$, which is the same
for all $k$ modes investigated.  }
\end{figure*}

\clearpage
\begin{figure*}
\begin{tabular}{c}
\psfig{file=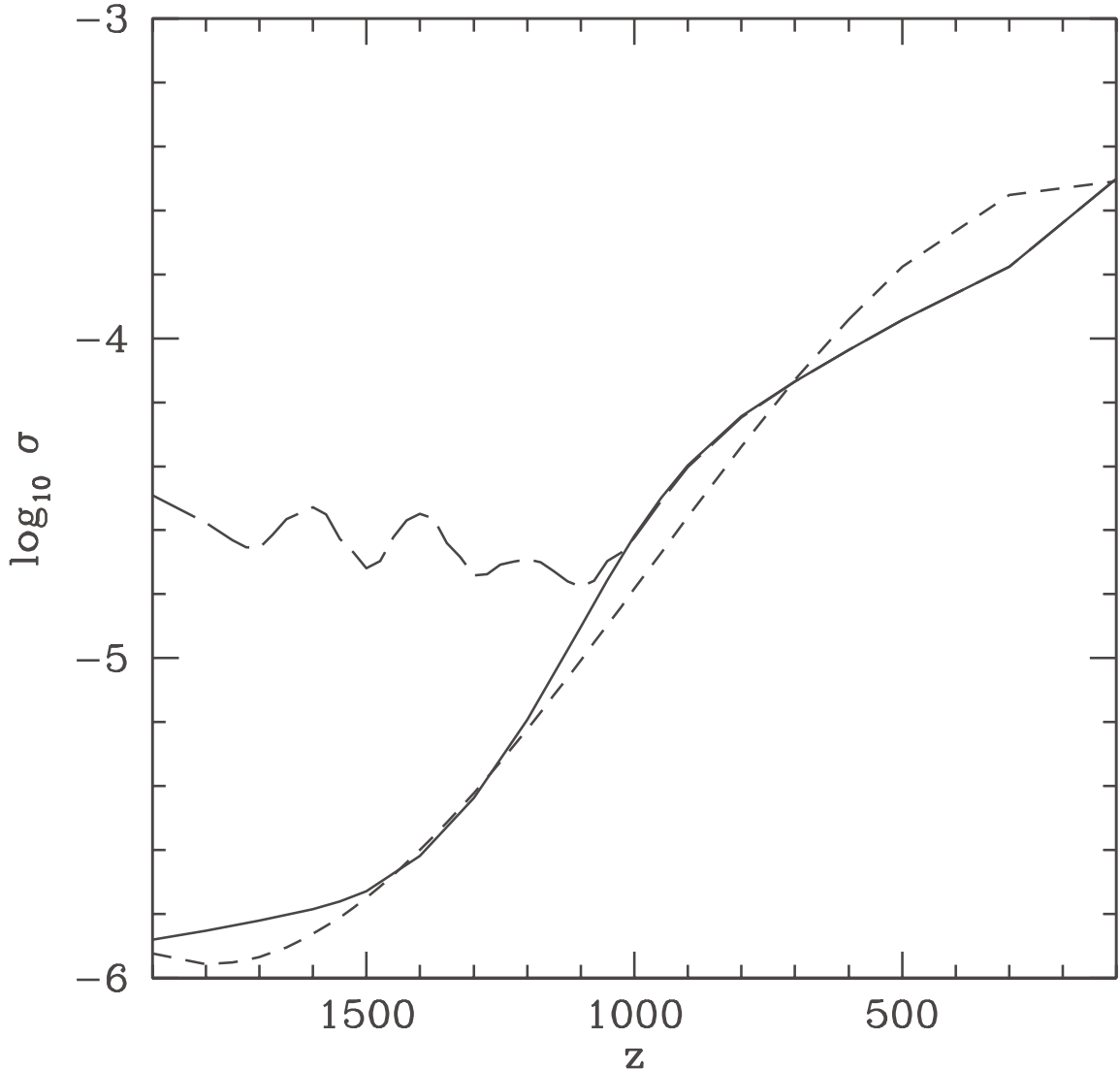,width=6in} 
\end{tabular}
\figurenum{4}
\caption{Evolution of $\sigma_v$, the rms of the baryon velocity $v_b$
(long dashed curve) and the rms of the baryon and photon velocity
difference $(v_b-v_\gamma)$ (solid curve).  The velocity difference is
small at $z\ga 1500$ due to the tight coupling but it increases
rapidly during and after recombination when baryons and photons are no
longer tightly coupled.  The baryons subsequently fall into the
potential well of the CDM and results in an increase in $v_b$ and
$(v_b-v_\gamma)$, whereas the photon perturbations simply diffuse
away.  The short-dashed curve shows the fitting formula for
$(v_b-v_\gamma)$ used in our calculations.}
\end{figure*}

\clearpage
\begin{figure*}
\begin{tabular}{c}
\psfig{file=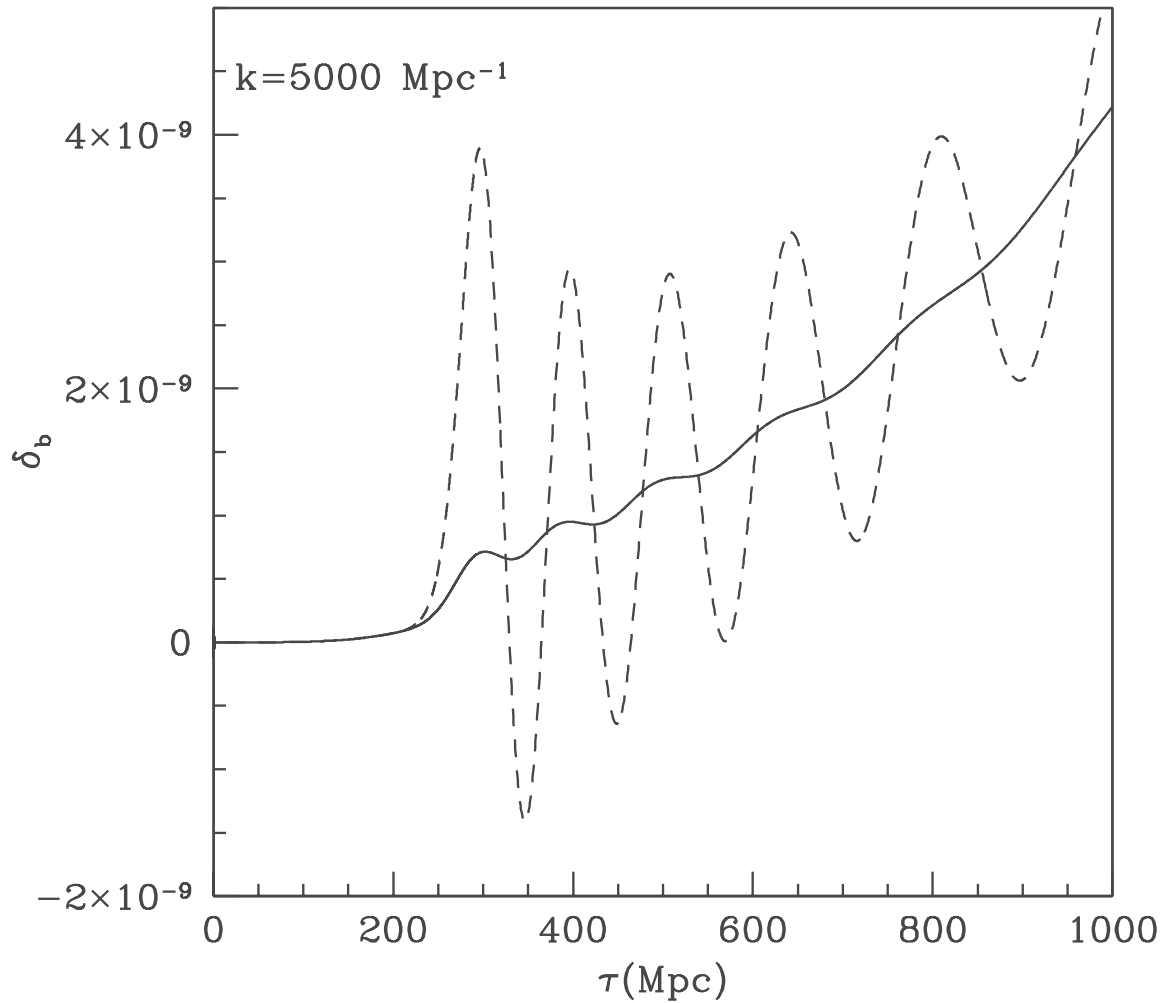,width=6in} 
\end{tabular}
\figurenum{5}
\caption{The linear (solid curve) vs. nonlinear (dashed curve) baryon
density field $\delta_{b}$ as a function of the conformal time $\tau$
for the $k=5000$ Mpc$^{-1}$ mode.  The Jeans oscillations discussed in
the text are clearly seen.  The effect of the second-order terms is to
increase the oscillation amplitude as well as shift slightly the
position of the peaks.  The oscillation period is $\sim 90$ Mpc
immediately after recombination. The conformal time is $\tau \approx
200$ Mpc at recombination and $\tau_0 \approx 10^4$ Mpc today.  }
\end{figure*}
 
\clearpage
\begin{figure*}
\begin{tabular}{cc}
\psfig{file=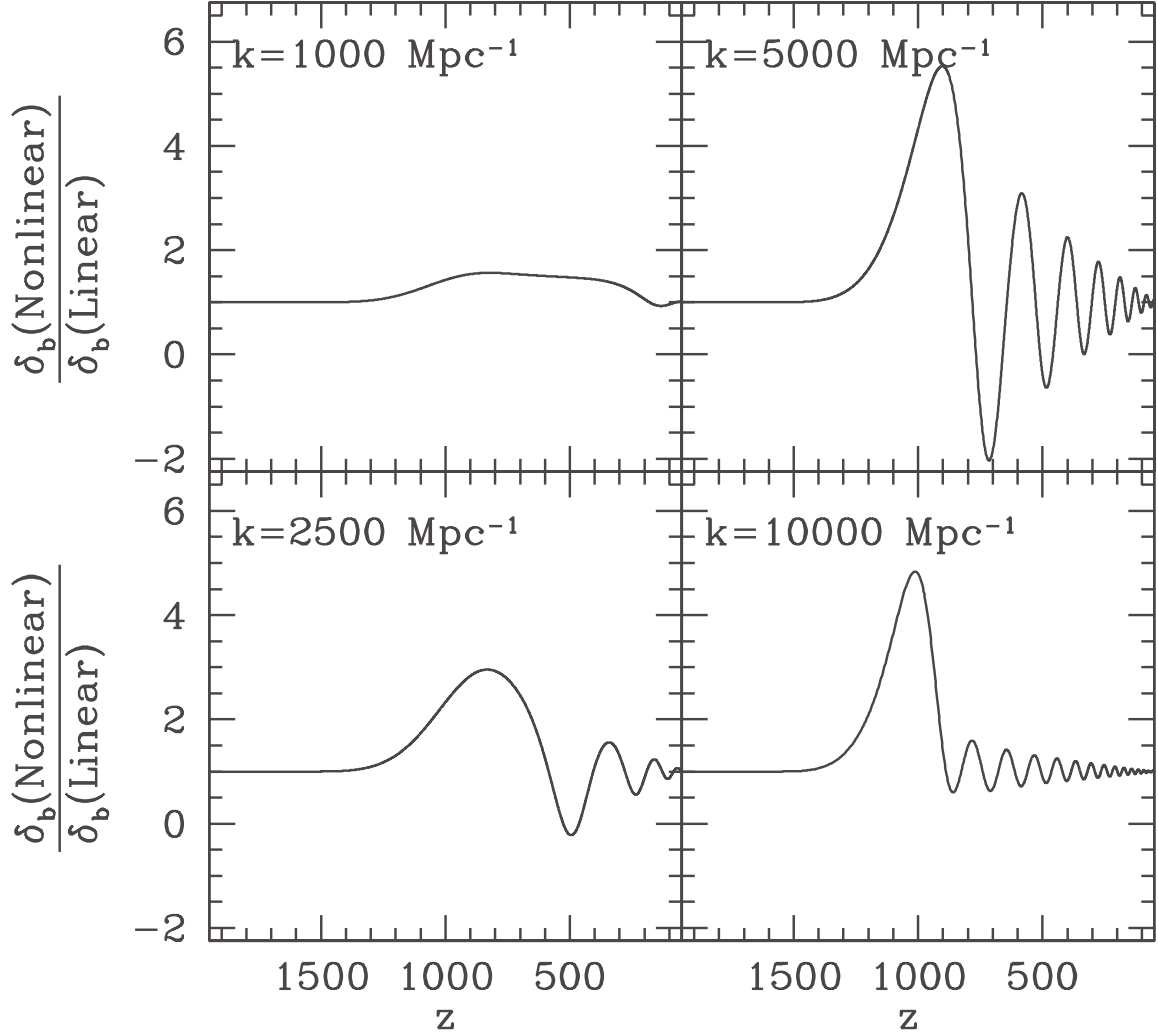,width=6in}
\end{tabular}
\figurenum{6}
\caption{Ratio of the nonlinear and linear baryon density field
$\delta_b$ for four $k$ modes: 1000, 2500, 5000, and 10000 Mpc$^{-1}$.
The oscillations in the ratio are more rapid for higher $k$ modes due
to the faster Jeans oscillations.  The maximum amplitude reached by a
particular $k$ mode is not a monotonic function of $k$ but peaks
between $k=5000$ and 10000 Mpc$^{-1}$. The oscillations die out and
the effect of the nonlinear term decreases at low redshifts when other
terms begin to dominate. }
\end{figure*}

\clearpage
\begin{figure*}
\begin{tabular}{cc}
\psfig{file=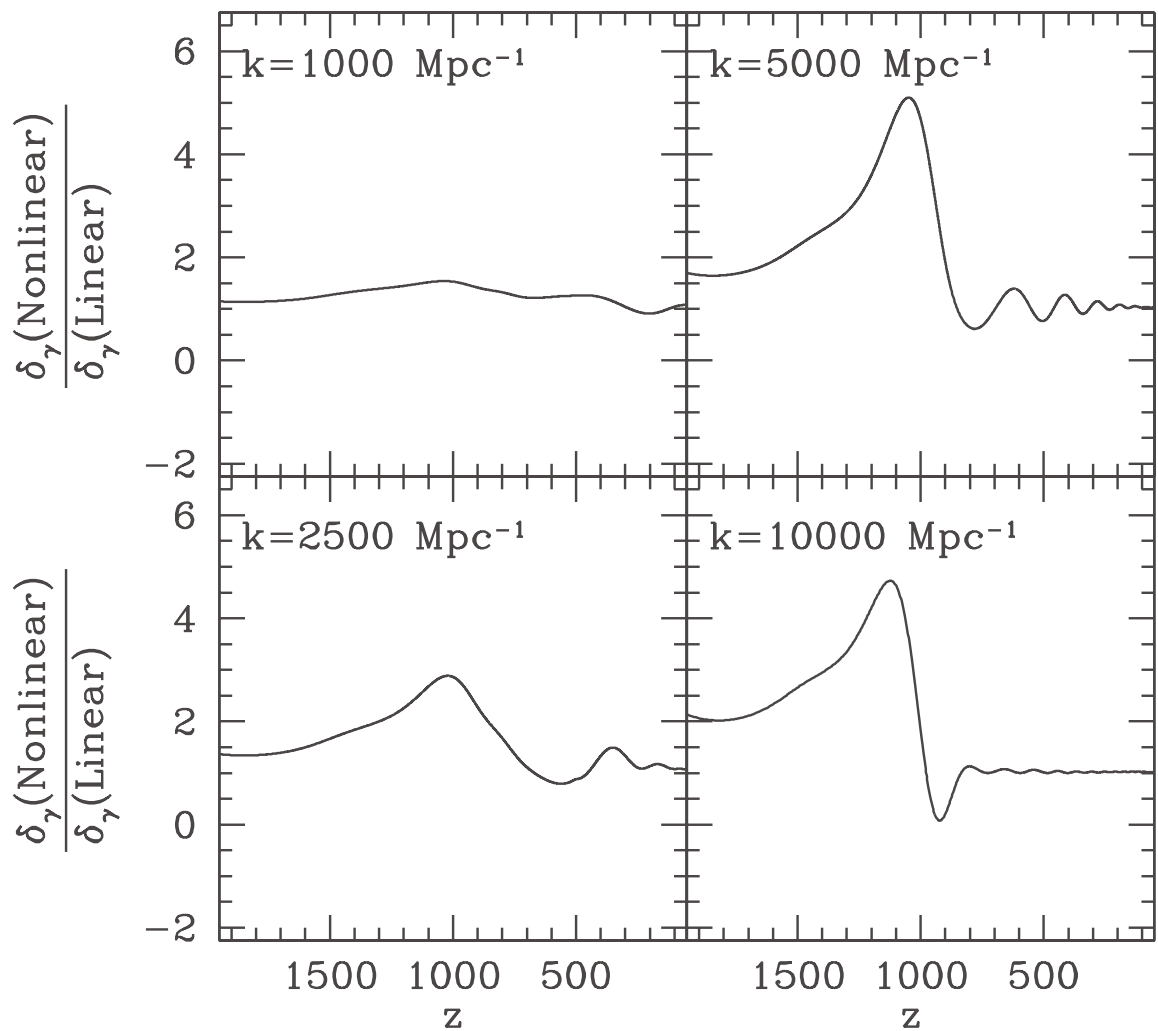,width=6in}
\end{tabular}
\figurenum{7}
\caption{Same as Fig.~6 but for the photon density field
$\delta_\gamma$.  The behavior is qualitatively similar to the baryons
in Fig.~6.  
}
\end{figure*}

\clearpage
\begin{figure*}
\begin{tabular}{cc}
\psfig{file=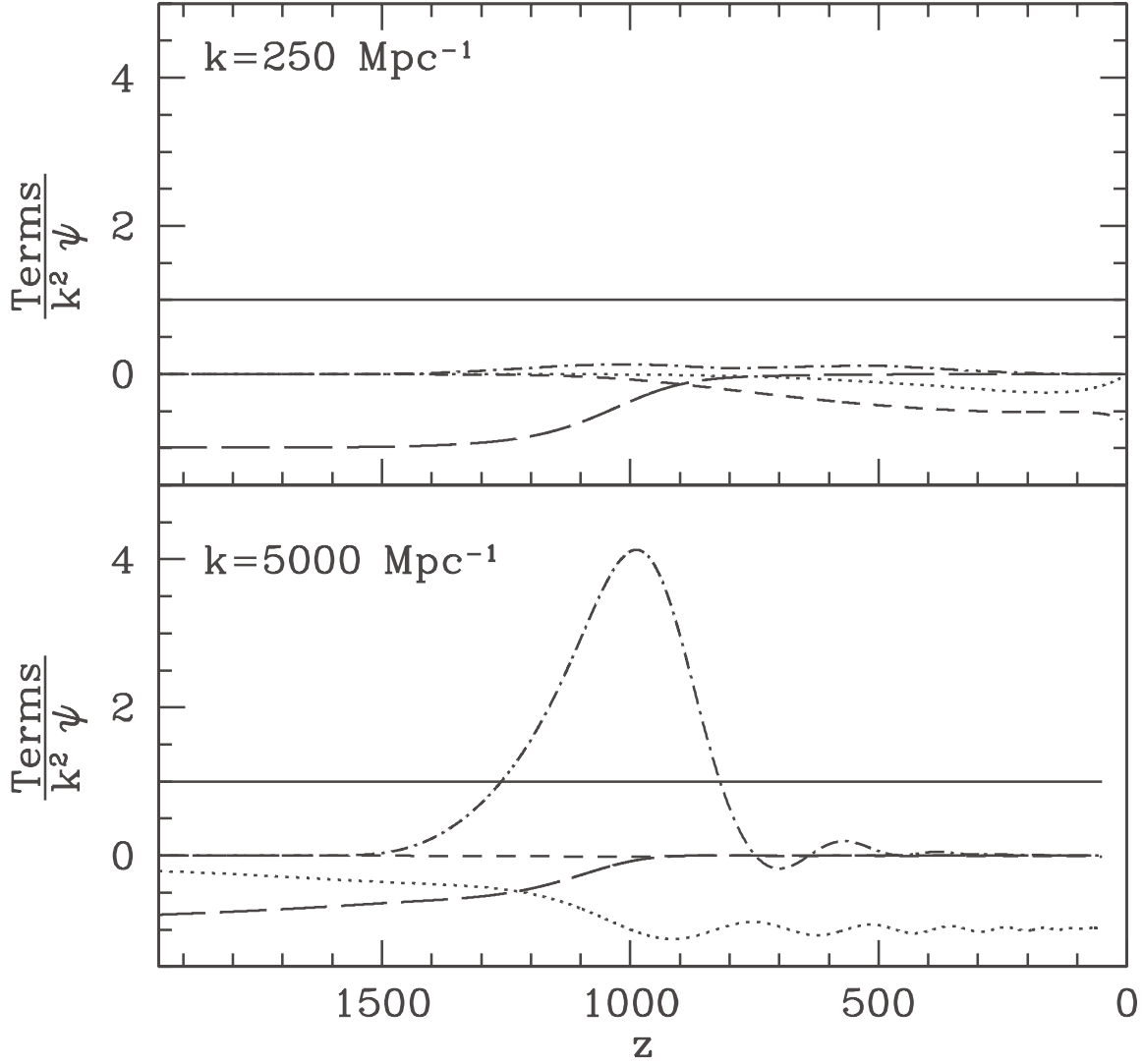,width=6in}
\end{tabular}
\figurenum{8}
\caption{ Amplitude of the individual terms on the right-hand side of
the baryon velocity equation~(\ref{theta}) relative to $k^2\psi$:
$-\dot{a}\theta_b/a$ (short-dashed), $c_s^2 k^2 \delta_b$ (dotted),
$(4 \rhogbar/ 3 \rhobbar) a \nebar \sigma_T (\theta_\gamma-\theta_b)$
(long dashed), and the second-order term (dot-dashed).  For the lower
$k$ mode, the second-order term is never important compared with the
first-order terms.  For $k=5000$ Mpc$^{-1}$, however, the near
cancellation among the first-order terms at $z\sim 1000$ allows the
second order term to dominate, leading to the nonlinear enhancement
shown in Figs.~6 and 7.  This figure illustrates how the second order
term becomes important for high $k$ modes.
}
\end{figure*}


\end{document}